\documentclass[10pt,conference]{IEEEtran}

\usepackage{amsmath,amssymb,amsfonts}
\usepackage{algorithmic}
\usepackage{graphicx}
\usepackage{textcomp}
\usepackage[table]{xcolor}
\usepackage{physics}
\usepackage{url}
\usepackage{tabularray}
\usepackage{multirow}
\usepackage{float}
\def\BibTeX{{\rm B\kern-.05em{\sc i\kern-.025em b}\kern-.08em
    T\kern-.1667em\lower.7ex\hbox{E}\kern-.125emX}}

\usepackage[ruled]{algorithm2e}
\usepackage{booktabs}
\usepackage{mathtools}
\usepackage{comment}

\usepackage[compact]{titlesec}
\titlespacing*{\section}{0pt}{1.6ex plus 0.4ex minus 0.2ex}{1.0ex plus .2ex minus .2ex}
\titlespacing*{\subsection}{0pt}{1.2ex plus 0.6ex minus 0.1ex}{.9ex plus .3ex minus .1ex}

\usepackage{cite}

\usepackage[hidelinks]{hyperref}
\hypersetup{colorlinks=true, citecolor=blue, linkcolor=blue, urlcolor=blue}
\usepackage{cleveref}

\makeatletter
\renewenvironment{abstract}{%
  \normalsize\noindent\textbf{Abstract}\par\noindent\ignorespaces
}{%
  \par
}
\renewenvironment{IEEEkeywords}{%
  \normalsize\noindent\textbf{Keywords}\par\noindent\ignorespaces
}{%
  \par
}
\makeatother

\definecolor{RED}{RGB}{190, 30, 49}
\definecolor{BLUE}{RGB}{11, 78, 179}
\definecolor{GREEN}{RGB}{0, 171, 86}
\definecolor{OLIVE}{RGB}{93, 150, 72}
\definecolor{GREY}{RGB}{50, 50, 50}
\definecolor{BROWN}{RGB}{122, 90, 40}

\newif\ifshowrevisions
\showrevisionsfalse
\newcommand{\rev}[1]{\ifshowrevisions{\color{BLUE}#1}\else#1\fi}

\newcommand{\be}{\begin{equation}}
\newcommand{\ee}{\end{equation}}

\newcommand{\bes}{\begin{equation*}}
\newcommand{\ees}{\end{equation*}}

\begin{document}

\title{
Extreme-Scale Linear-Scaling Kohn-Sham DFT at 100 Million Atoms: Bridging Quantum Simulations and Experiments
}

\author{
\begin{minipage}{0.98\textwidth}
\centering
\normalsize
Qimen Xu\textsuperscript{1,*,$\dagger$},
Yu Zhang\textsuperscript{1,*},
Dixing Ni\textsuperscript{1},
Lei Gao\textsuperscript{2},
Guangnan Feng\textsuperscript{3},
Qinrui Zheng\textsuperscript{1},
Jianting Liu\textsuperscript{1}\\
Haitian Lu\textsuperscript{4},
Zhaopeng Jia\textsuperscript{5},
Wei Xue\textsuperscript{5},
Shriram Chandran\textsuperscript{6},
Torsten Hoefler\textsuperscript{6,7,$\dagger$},
Haohuan Fu\textsuperscript{1,8,$\dagger$},
Yutong Lu\textsuperscript{1,3,$\dagger$}
\par\vspace{0.6em}
\small
\textsuperscript{1}National Supercomputing Center in Shenzhen, Guangdong, China\\
\textsuperscript{2}Peking University Shenzhen Graduate School, Guangdong, China\\
\textsuperscript{3}Sun Yat-sen University, Guangdong, China\\
\textsuperscript{4}National Supercomputing Center in Wuxi, Jiangsu, China\\
\textsuperscript{5}Tsinghua University, Beijing, China\\
\textsuperscript{6}ETH Zurich, Zurich, Switzerland\\
\textsuperscript{7}ADIA Lab Fellow\\
\textsuperscript{8}Tsinghua Shenzhen International Graduate School, Guangdong, China
\par\vspace{0.5em}
\textsuperscript{*}These authors contributed equally to this work.\\
\textsuperscript{$\dagger$}Corresponding authors:
xuqm@nsccsz.cn;
torsten.hoefler@inf.ethz.ch;\\
haohuan@tsinghua.edu.cn;
luyutong@mail.sysu.edu.cn
\end{minipage}
}

\maketitle

\vspace*{-5pt}
\begin{abstract}
Kohn-Sham density functional theory (DFT) remains the workhorse of
\emph{ab initio} materials simulation, yet cubic computational and quadratic 
memory scaling have confined calculations to a few hundred to thousands 
of atoms, spanning only nanometers, far below experimentally relevant 
length scales. We introduce XLSDFT, a linear-scaling DFT framework based
on divide-and-conquer decomposition of the one-particle density matrix 
and Chebyshev-filtered subspace iteration, achieving linear computational 
and memory scaling while retaining DFT accuracy. Deployed on the LineShine 
exascale supercomputer, XLSDFT reduces computational complexity by orders 
of magnitude, enabling unprecedented DFT scale: 
\rev{a 200-million-atom silicon crystal, twentyfold beyond the prior
record.} Our implementation achieves \rev{96.6\% weak-scaling efficiency and
sustained 157.9~Pflop/s (FP64) for a 100-million-atom scaling study.} We further
simulate an 11-million-atom all-solid-state battery interface 
of unprecedented complexity, $1{,}000$ times beyond prior DFT for such systems, revealing how
lithium metal reacts with the solid electrolyte at atomic resolution, in
quantitative agreement with spectroscopy experiments.
\end{abstract}


\section{Contributions and Performance Highlights}
Two world records for Kohn-Sham DFT:
\rev{(1)~100- and 200-million-atom silicon, an order of magnitude} beyond the prior
10-million-atom record; the 100-million-atom study achieved
\rev{157.9~FP64~Pflop/s sustained, 1.31~FP64~Eflop/s peak in subspace rotation, and
96.6\% weak-scaling efficiency} on the LineShine system;
(2)~an 11-million-atom all-solid-state battery interface,
$1,000\times$ beyond \rev{DFT SoTA for such complex interfacial systems.}


\section{Computational Performance Summary}
\vspace{-8pt}
\begin{table}[H]
\centering
\caption{Summary of Performance Attributes}
\label{tab:performance-attributes}
\begin{tabular}{@{}ll@{}}
    \toprule
    Performance attribute & Reported result \\
    \midrule
    Category of achievement & Scalability, time-to-solution, peak performance \\
    Type of method used & Linear-scaling DFT, D\&C, CheFSI \\
    Results reported based on & Full application timing; timed kernel phases \\
    Precision reported & Double precision \\
    System scale & Results measured on full-scale system \\
    Measurement mechanism & Timers, analytic flop counts, parallel efficiency \\
    \bottomrule
\end{tabular}
\end{table}


\section{Overview of the Problem}
\label{sec:overview}

Kohn-Sham (KS) density functional theory
(DFT)~\cite{hohenberg1964inhomogeneous,kohn1965self} is the
workhorse of \emph{ab initio} materials simulation, providing access
to ground-state electronic structure, including charge density, energy, density of
states, and atomic forces, with predictive accuracy
across a wide range of materials.
The fundamental obstacle to its wider application is computational
cost: the dominant step, iterative diagonalization of the KS
Hamiltonian, scales as $\mathcal{O}(N^3)$ in the number of atoms
$N$, confining routine calculations to a few hundred to a few thousand
atoms, corresponding to only a few nanometers in physical
size~\cite{GB2023,ls3df,deepmd_assb}.
On the other hand, the solid--solid interfaces and interphase layers
that govern battery performance, semiconductor device behavior, and
heterogeneous catalysis extend over length scales from many dozens of
nanometers to several micrometers, as directly characterized by
experimental probes such as X-ray photoelectron spectroscopy (XPS),
time-of-flight secondary-ion mass spectrometry (ToF-SIMS), and
related techniques.

\begin{figure}[htbp]
  \centering
  \includegraphics[width=\linewidth]{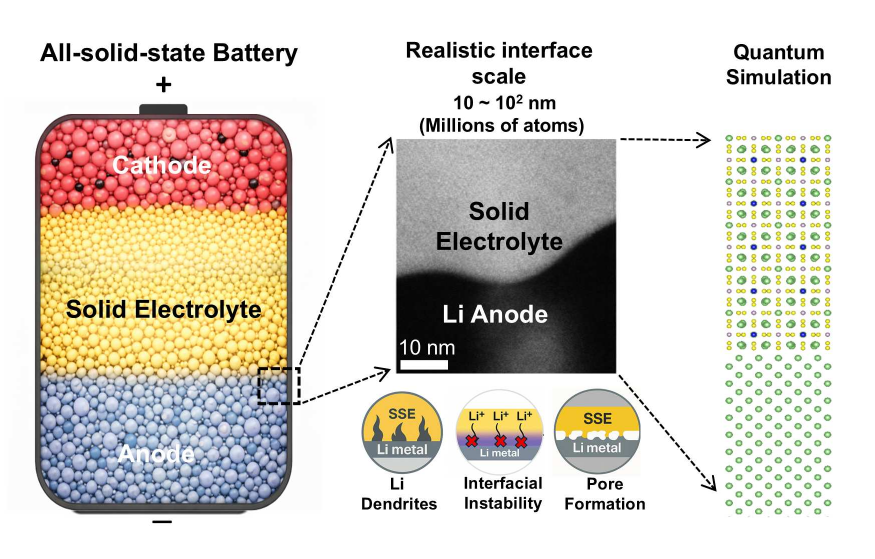}
  \caption{The length-scale barrier in \emph{ab initio} simulation for
    battery interfaces.
    Conventional KS-DFT is typically confined to the nanometer scale, whereas
    experimentally relevant interfacial regions extend to tens of nanometers
    and beyond, containing millions of atoms that are beyond the practical
    reach of cubic-scaling DFT.}
  \label{fig:length-scale}
\end{figure}

This mismatch has prevented direct, quantitative comparison
between DFT-computed electronic-structure observables and experimental
measurements for most real-world heterogeneous materials systems.
The resulting length-scale gap is summarized in Fig.~\ref{fig:length-scale}.
Despite decades of progress in linear-scaling DFT methods and recent
advances in machine-learning approaches to electronic structure, no
existing method can deliver DFT-level observables for the chemically
complex, multi-element interfacial systems whose physical features
span tens of nanometers to several micrometers;
a detailed review is provided in Sec.~\ref{sec:state_of_the_art}.

In this work, we present XLSDFT, an extreme-scale $\mathcal{O}(N)$ KS-DFT code
based on divide-and-conquer (D\&C) decomposition of the one-particle
density matrix and Chebyshev-filtered subspace iteration (CheFSI),
deployed on the LineShine exascale supercomputer at NSCC-SZ
(Sec.~\ref{sec:innovations}).
We demonstrate XLSDFT on two \rev{classes of systems: an
11-million-atom Li/LGPS solid-state battery interface ($57\times44\times90$~nm),
the first DFT simulation at the scale of experimentally characterized
interphase layers, with electronic-structure observables validated
against XPS measurements; and 100-million-atom and 200-million-atom silicon
crystals, an order of magnitude beyond the prior DFT world record.}

All-solid-state batteries promise superior energy density and safety over
conventional lithium-ion cells, but commercialization is blocked by interfacial
instability: LGPS, despite record room-temperature Li-ion conductivity, decomposes
reductively upon contact with Li metal, building a resistance layer that grows with
every cycle.
The atomic decomposition mechanism and the spatial distribution of reaction products
are experimentally observed but computationally inaccessible -- prior DFT is
confined to few-nanometer models, far below the tens-of-nanometer interphase
thickness where degradation actually occurs.

For the battery interface, XLSDFT serves as the electronic-structure
core of a three-stage multi-scale workflow:

\vspace{2pt}
\noindent\textbf{Stage~1: Structure generation via MLFF-MD.}
A DeePMD-kit machine-learning force field~\cite{deepmd}, trained on
\emph{ab initio} molecular dynamics (AIMD) trajectories of small
($\sim$100--500 atom) representative interface cells, is deployed for
molecular dynamics at the full $\sim$11-million-atom scale, generating
thermally equilibrated interface configurations at near-DFT accuracy.

\vspace{2pt}
\noindent\textbf{Stage~2: Electronic structure via XLSDFT.}
The MLFF-generated atomic structure is passed to XLSDFT, which
performs a full KS-DFT calculation at $\mathcal{O}(N)$ cost,
computing the electron density, density of states (DOS), projected
DOS (PDOS), Bader charges, and electrostatic potential natively
without additional overhead.

\vspace{2pt}
\noindent\textbf{Stage~3: Experimental cross-validation.}
Computed observables are compared
quantitatively against XPS measurements on the same
material system, providing the first direct DFT--experiment comparison
at the scale of physically representative interfacial models.


\section{Current State of the Art}
\label{sec:state_of_the_art}

\subsection{Large-Scale \emph{Ab Initio} Simulation: Progress and Limitations}
\label{subsec:reduced_scaling}

Many efforts have been devoted to modeling large-scale materials from
first principles, with key milestones summarized in
Table~\ref{tab:sota}.
Despite progressive gains in accessible system size, every existing
approach faces fundamental limitations that prevent application to
chemically complex interfacial systems at experimentally relevant scales.

\begin{table*}[htb]
  \centering
  \caption{Performance milestones in large-scale KS-DFT.
    Performance rates are in \textnormal{Pflop/s} and use FP64 unless noted.
    Wall times are per SCF iteration.}
  \label{tab:sota}
  \small
  \setlength{\tabcolsep}{3.5pt}
  \renewcommand{\arraystretch}{1.22}
  \begin{tabular}{@{}llllll r r r@{}}
    \toprule
    Code & Year & Basis & Electronic solver & System & Machine & Atoms & min/SCF & Pflop/s \\
    \midrule
    QBox~\cite{qbox}        & 2006 & PW    & Steepest descent                         & Mo                   & BlueGene/L & 1k   & 8.8      & 0.2  \\
    RSDFT~\cite{rsdft}      & 2011 & FD    & CG + subspace diag.                   & Si nanowire          & K computer & 107k & 73.6     & 7.1  \\
    DFT-FE~\cite{dftfe}     & 2019 & FE    & CheFSI   & Mg dislocation       & Summit     & 11k  & 2.4      & 46   \\
    CONQUEST~\cite{conquest} & 2020 & PAO  & LNV DM minim. (LS)                    & bulk Si              & K computer & 1M   & ---      & ---  \\
    DGDFT~\cite{dgdft}      & 2022 & DG-PW & PEXSI   & Li/Na                 & Sunway     & 2.5M & ---      & 64   \\
    DFT-FE~\cite{GB2023}    & 2023 & FE    & CheFSI   & Mg-Y alloy           & Frontier   & 74k  & 8.6      & 660  \\
    PARSEC~\cite{dogan2023solving} & 2023 & FD & CheFSI & Si/H nanocluster & Frontera & 117k & 200.6    & --- \\
    PARSEC~\cite{dogan2023real}    & 2023 & FD & CheFSI & Si/H nanocluster & Frontier & 216k & $\sim$300 & --- \\

    LS3DF~\cite{ls3df}      & 2024 & PW    & Frag. LS + AB-CG                      & bulk Si              & Sugon      & 10M  & $\sim$ 2.2 & 34.8 \\
    \midrule
    \multirow{3}{*}{\textbf{XLSDFT (this work)}}
      & \multirow{3}{*}{\textbf{2026}}
      & \multirow{3}{*}{\textbf{FD}}
      & \multirow{3}{*}{\textbf{LS+CheFSI}}
      & \textbf{Li/LGPS}  & \textbf{LineShine} & \textbf{11M}  & \rev{0.6} & \rev{96.8} \\
    & & & & \textbf{Si}       & \textbf{LineShine} & \textbf{100M} & \rev{1.2} & \rev{157.9} \\
    & & & & \rev{\textbf{Si}} & \rev{\textbf{LineShine}} & \rev{\textbf{200M}} & \rev{2.4} & \rev{157.6} \\
    \bottomrule
  \end{tabular}
\end{table*}

Two broad strategies have been pursued to extend the reach of KS-DFT.
The first focuses on basis sets and discretization schemes.
The plane-wave (PW) basis offers systematic convergence through spectral
accuracy and underpins widely used codes such as VASP~\cite{vasp},
Quantum ESPRESSO~\cite{qe}, and QBox~\cite{qbox}, but its
globally delocalized nature constrains parallel scalability as system size
grows.
Localized basis sets alleviate communication costs: Gaussian-type orbitals
in CP2K~\cite{cp2k} and numerical atomic orbitals in
SIESTA~\cite{siesta} reduce data movement but lack systematic
improvability, introducing basis-set error that is difficult to control for
metallic or chemically complex systems.
Systematically convergent real-space discretizations avoid both drawbacks:
finite-difference (FD) methods, as implemented in RSDFT~\cite{rsdft},
PARSEC~\cite{kronik2006parsec}, 
and SPARC~\cite{xu2021sparc}, and
finite-element (FE) methods in DFT-FE~\cite{dftfe}, both achieve
favorable parallel scalability on large GPU-based platforms while permitting
systematic basis refinement.
Yet all of these codes retain $\mathcal{O}(N^3)$ scaling and in routine
practice are limited to $\sim$1,000 atoms.
The largest systems reported with cubic-scaling codes are 74,000 atoms with
DFT-FE~\cite{GB2023} and 200,000 atoms with PARSEC~\cite{dogan2023real},
both requiring dedicated leadership-class resources.
The second strategy exploits the \emph{nearsightedness principle}~\cite{prodan2005nearsightedness}:
the one-particle density matrix decays exponentially in real space, so
electronic observables can be assembled from local information, enabling
sub-cubic or linear-scaling algorithms.
Sub-linear-scaling methods such as DGDFT~\cite{dgdft} exploit quasi-2D
geometry via a discontinuous Galerkin basis with PEXSI, achieving
$\mathcal{O}(N^{1.5})$ effective scaling and reaching 2.5~million atoms
for metallic heterostructures at 64~Pflop/s, the prior sustained-performance
high-watermark for ground-state DFT.
This advantage is, however, tied to quasi-2D geometry and does not
generalize to three-dimensional heterogeneous interfaces.
True $\mathcal{O}(N)$ methods include ONETEP~\cite{onetep} and
CONQUEST~\cite{conquest} (localized orbital bases; up to
$\sim$1~million-atom Si), and SQDFT~\cite{sqdft} (spectral
quadrature; efficient at high temperature, but carries a large
prefactor at moderate temperatures that grows steeply as the mesh is
refined; a 1-million-atom Al simulation at room
temperature was demonstrated in~\cite{gavini2023roadmap}, made feasible by
Al's soft pseudopotential allowing a coarse mesh).
Most recently, LS3DF~\cite{ls3df} reached 10~million atoms for a silicon
crystal at 34.8~Pflop/s, the prior DFT world record.
\rev{XLSDFT extends this record to a 100-million-atom and a 200-million-atom silicon
calculation.}
All of these methods have been demonstrated only on single-element or
single-phase insulating and semiconducting systems; none supports the
full suite of electronic-structure observables for a metallic
multi-element interface.

Machine-learning force fields (MLFFs) such as DeePMD-kit~\cite{deepmd}
provide an orthogonal route, achieving near-DFT accuracy for atomic
forces and energies at millions of atoms.
The largest MLFF study of an ASSB interface~\cite{deepmd_assb} reached
$\sim$12,000 atoms but yields no electronic-structure observables: band
alignment, charge transfer, and projected DOS are entirely inaccessible.
Semiempirical tight-binding methods (e.g., GFN-xTB in
CP2K~\cite{nolsm}) can reach 100~million atoms and larger at mixed precision
but sacrifice self-consistent accuracy.
\subsection{Machine-Learning Electronic Structure Methods}

Recent efforts have sought to predict electronic-structure quantities
directly from atomic structure using machine learning.
DeepH~\cite{deepH} predicts DFT Hamiltonian matrix elements using
equivariant neural networks, enabling approximate band structures at
reduced cost.
ChargE3Net~\cite{charge3net} and a super-resolution-inspired
approach~\cite{sr_density} predict the electron density directly from
atomic coordinates.
These are promising directions, but none yet meets the requirements
for production electronic-structure analysis of large, complex systems.
First, all methods have been trained and validated on systems of fewer
than a few hundred atoms in simple or single-component chemical
environments; their accuracy for chemically complex multi-element
interfacial systems is unestablished.
Second, even if the electron density is accurately predicted, the
KS eigenstates must still be computed via a
DFT solve to obtain orbital-resolved observables such as the
density of states, projected DOS, and Bader charges.
Third, none of these methods has been demonstrated on systems larger
than a few nanometers; scaling to the tens-of-nanometer interfacial
models required here is an open problem.
At present, no machine-learning method can replace KS-DFT for
electronic-structure analysis of large, chemically complex systems.

\subsection{The Gap This Work Addresses}

The combination of the $\mathcal{O}(N^3)$ wall of conventional DFT,
the limitations of existing $\mathcal{O}(N)$ methods for metallic
multi-element systems, the inability of MLFFs to provide
electronic-structure observables, and the immaturity of ML
electronic-structure methods at scale defines the gap this work
addresses.
XLSDFT closes it through two complementary advances: reducing
algorithmic complexity from $\mathcal{O}(N^3)$ to $\mathcal{O}(N)$,
and executing the remaining operations at high hardware efficiency
via exascale-optimized parallelism and architecture-specific kernels
(ARM SVE/SME, \rev{custom dense linear algebra, and HBM-resident
memory-intensive operations}).
Together, these enable production-quality KS-DFT delivering the full
suite of electronic-structure observables for chemically complex
interfacial systems at experimentally relevant length scales,
with FP64 dominant arithmetic and targeted use of FP32/FP16 only in limited roles (AMG preconditioning and orbital warm-start storage).


\section{Innovations Realized}
\label{sec:innovations}

\begin{figure}[!b]
  \centering
  \includegraphics[width=\columnwidth,trim=0 0 70pt 0,clip]{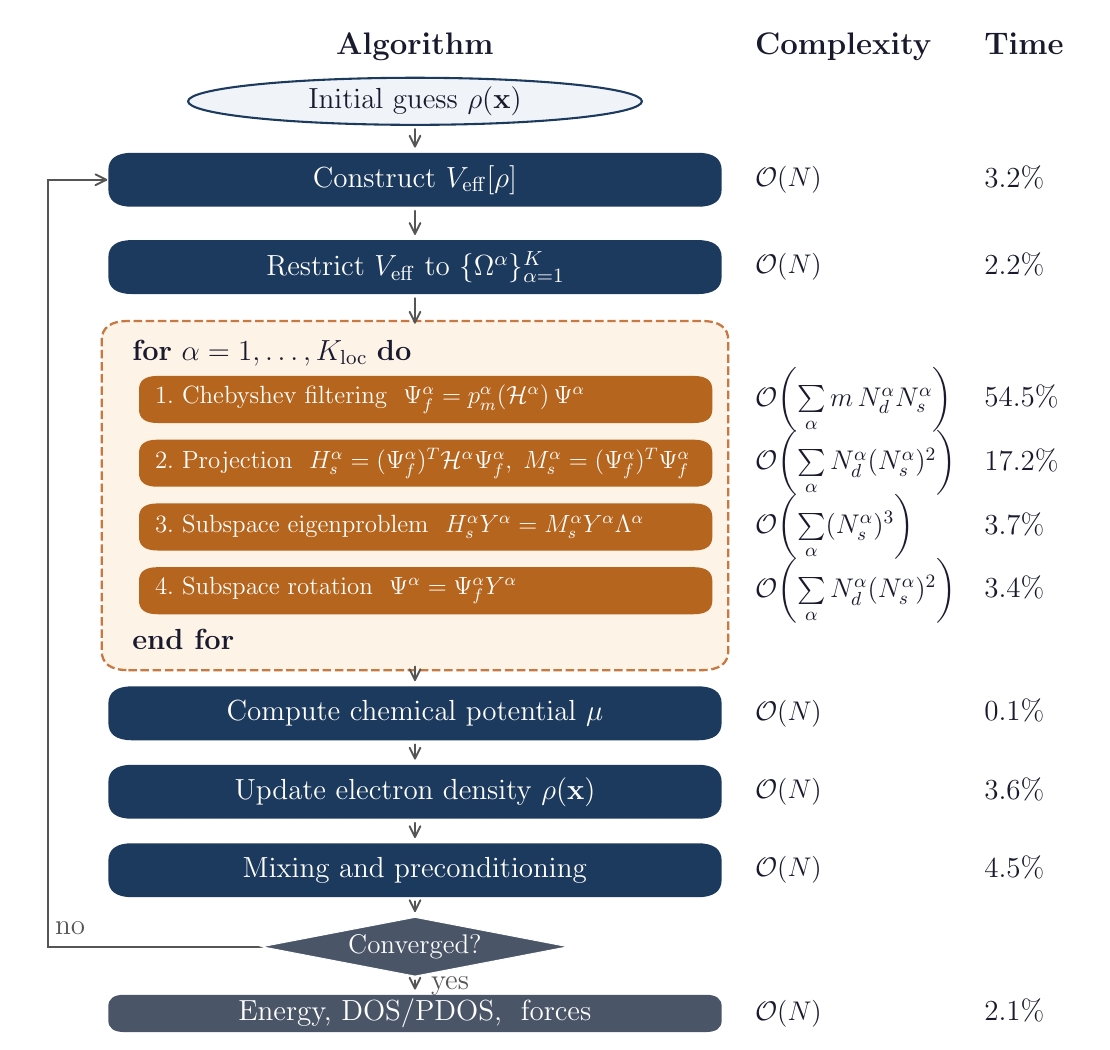}
  \caption{%
    Self-consistent field (SCF) flowchart for XLSDFT.
    \rev{Each row shows the algorithmic step (left) and its asymptotic cost per
    SCF iteration (right).}
    All phases scale as $\mathcal{O}(N)$ or better.
    ($K$: total number of subsystems; $K_\text{loc}$: subsystems assigned per MPI process.)}
  \label{fig:xlsdft-overview}
\end{figure}

The electronic ground state of a system of $N$ atoms and $N_e$ valence
electrons is determined by the KS
eigenproblem~\cite{hohenberg1964inhomogeneous,kohn1965self}
(neglecting spin and restricting to $\Gamma$-point sampling):
\begin{equation}
  \left(\mathcal{H} \equiv -\tfrac{1}{2}\nabla^2 + V_\mathrm{xc} + \phi +
    V_\mathrm{nl}\right)\,\psi_n = \lambda_n\psi_n,
  \quad n = 1,\ldots,N_s,
  \label{eq:ks}
\end{equation}
where $V_\mathrm{xc}$, $\phi$, and $V_\mathrm{nl}$ are the
exchange-correlation, electrostatic, and nonlocal pseudopotential
operators, respectively; $\psi_n$ and $\lambda_n$ are the KS orbitals
and eigenvalues; and $N_s \approx N_e/2$ is the number of occupied
states.
The dominant computational step, iterative diagonalization of
$\mathcal{H}$ over $N_d$ degrees of freedom to find the $N_s$
occupied orbitals, costs $\mathcal{O}(N_d N_s^2)$,
yielding $\mathcal{O}(N^3)$ overall cost.
XLSDFT overcomes this wall through the algorithmic and HPC innovations
described below (summarized in Fig.~\ref{fig:xlsdft-overview}).

\subsection{XLSDFT: Linear-Scaling DFT Algorithm}

\subsubsection{Density Matrix Formulation and Nearsightedness}

To overcome the $\mathcal{O}(N^3)$ scaling bottleneck, XLSDFT is
formulated in terms of the \emph{one-particle density
matrix}~\cite{goedecker1999}
\begin{equation}
  \mathcal{D} = g(\mathcal{H};\mu,\sigma)
    = \left(1+\exp\!\left(\tfrac{\mathcal{H}-\mu\mathcal{I}}{\sigma}
      \right)\right)^{-1},
  \label{eq:dm}
\end{equation}
where $g(\lambda;\mu,\sigma)$ is the Fermi-Dirac function,
$\mu$ is the chemical potential, and $\sigma = k_BT$ is the
electronic smearing.
All ground-state observables follow from $\mathcal{D}$: notably,
the electron density is the diagonal
$\rho(\mathbf{x}) = 2\mathcal{D}(\mathbf{x},\mathbf{x})$.
The electrostatic potential $\phi$ is obtained by solving the
Poisson equation~\cite{ghosh2016higher,ghosh2017sparc}
\begin{equation}
  -\tfrac{1}{4\pi}\nabla^2\phi = \rho + b,
  \label{eq:poisson}
\end{equation}
where $b = \sum_I b_I$ is the total pseudocharge density,
$b_I = -(1/4\pi)\nabla^2 V_I$, and $V_I$ is the local
pseudopotential of atom $I$.
This local reformulation of electrostatics~\cite{pask2005real}
replaces the $\mathcal{O}(N^2)$ Hartree energy summation with an
$\mathcal{O}(N)$ Poisson solve.

The key observation enabling $\mathcal{O}(N)$ cost is the
\emph{nearsightedness principle}~\cite{prodan2005nearsightedness}:
for insulating as well as metallic systems at finite 
temperature~\cite{goedecker1998decay,benzi2013decay}, the
density matrix decays exponentially in real space,
$|\mathcal{D}(\mathbf{x},\mathbf{x}')| \lesssim
\exp(-|\mathbf{x}-\mathbf{x}'|/\xi)$, whereas the individual KS
orbitals $\psi_n$ may be fully delocalized.
This locality of $\mathcal{D}$ is the theoretical foundation of the
divide-and-conquer construction described next.

\subsubsection{Divide-and-Conquer Domain Partitioning}

In XLSDFT, we start from the spectral representation of the global
density matrix:
\begin{equation}
  \mathcal{D}(\mathbf{x},\mathbf{x}') = \sum_n
  g(\lambda_n;\mu,\sigma)\,\psi_n(\mathbf{x})\psi_n(\mathbf{x}'),
  \label{eq:dm_global}
\end{equation}
where $g(\lambda;\mu,\sigma) = (1+\exp((\lambda-\mu)/\sigma))^{-1}$
is the Fermi-Dirac occupation with chemical potential $\mu$ and
smearing $\sigma = k_B T$.
All electronic observables of interest follow from
$\mathcal{D}$: the electron density is
$\rho(\mathbf{x}) = 2\mathcal{D}(\mathbf{x},\mathbf{x})$,
i.e., $\rho(\mathbf{x}) = 2\sum_n g_n|\psi_n(\mathbf{x})|^2$.
Computing Eq.~\eqref{eq:dm_global} globally requires all $N_s \sim N$
eigenpairs of $\mathcal{H}$, which is the $\mathcal{O}(N^3)$ bottleneck.

Nearsightedness justifies replacing the global density matrix with
a sum of \emph{local} subsystem density matrices, each defined over
a spatially restricted subdomain.
We partition the global domain $\Omega$ into subdomains
$\{\Omega^\alpha\}$, each comprising an
\emph{interior region} $\Omega^\alpha_C$ and a surrounding
\emph{buffer region} of thickness $R_\text{buffer}$:
\begin{equation}
  \begin{gathered}
  \Omega^\alpha = \Omega^\alpha_C \cup \Omega^\alpha_\text{buffer}, \quad
  \Omega = \bigcup_\alpha \Omega^\alpha_C, \\
  \Omega^\alpha_C \cap \Omega^{\alpha'}_C = \varnothing
  \;\; (\alpha \neq \alpha').
  \end{gathered}
\end{equation}

\begin{figure}[htbp]
  \centering
  \includegraphics[width=0.7\columnwidth]{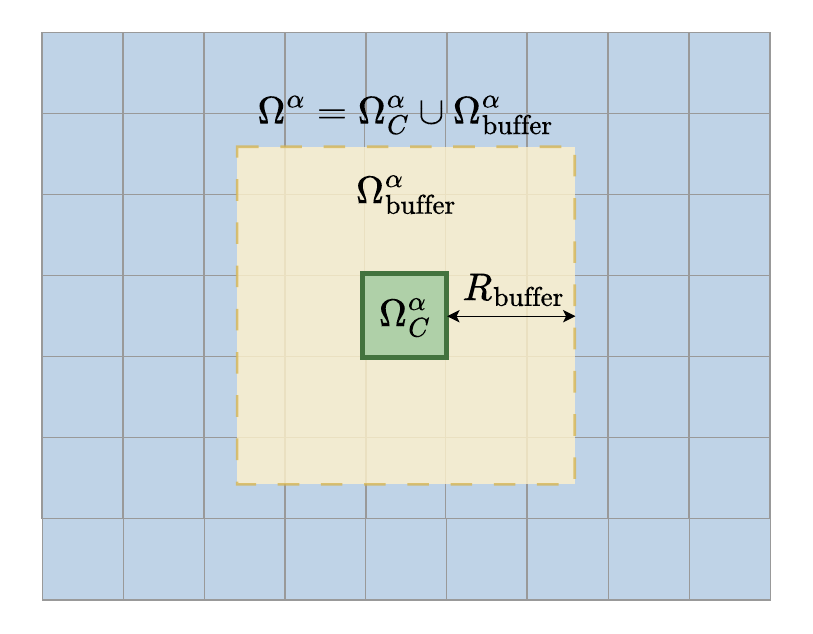}
  \caption{Divide-and-conquer domain partitioning in XLSDFT.
    The simulation cell $\Omega$ is partitioned into non-overlapping
    core subdomains $\{\Omega^\alpha_C\}$; each core is extended by a
    buffer region $\Omega^\alpha_\text{buffer}$ of thickness
    $R_\text{buffer}$ to form the subdomain $\Omega^\alpha$ on which
    the local Hamiltonian $\mathcal{H}^\alpha$ is solved independently.}
  \label{fig:domain-partition}
\end{figure}
For each subdomain, the global effective potential $V_\text{eff}$
is restricted to $\Omega^\alpha$ to define the local Hamiltonian
$\mathcal{H}^\alpha$, whose eigenpairs
$\{\lambda_n^\alpha,\psi_n^\alpha\}$ satisfy
$\mathcal{H}^\alpha\psi_n^\alpha = \lambda_n^\alpha\psi_n^\alpha$.
The subsystem density matrix then admits the spectral representation
\begin{equation}
  \mathcal{D}^\alpha(\mathbf{x},\mathbf{x}') = \sum_n
  g(\lambda_n^\alpha;\mu,\sigma)\,
  \psi_n^\alpha(\mathbf{x})\psi_n^\alpha(\mathbf{x}'),
\end{equation}
which has exactly the same form as Eq.~\eqref{eq:dm_global} but
involves only the $N_s^\alpha \ll N_s$ local eigenpairs of the
reduced-size Hamiltonian $\mathcal{H}^\alpha$.
The buffer region ensures that
$\mathcal{D}^\alpha(\mathbf{x},\mathbf{x}') \approx
\mathcal{D}(\mathbf{x},\mathbf{x}')$ for
$\mathbf{x}\in\Omega^\alpha_C$,
with exponential convergence as $R_\text{buffer}$
increases~\cite{yang1995density,suryanarayana2017on}.

The global density matrix is therefore approximated by restricting
each subsystem contribution to its interior region via a binary
partition of unity $\{p^\alpha\}$,
$\sum_\alpha p^\alpha = 1$, with
$p^\alpha(\mathbf{x}) = 1$ for $\mathbf{x}\in\Omega^\alpha_C$
and zero otherwise:
\begin{equation}
  \mathcal{D}(\mathbf{x},\mathbf{x}') \approx
  \sum_\alpha p^\alpha(\mathbf{x})\,\mathcal{D}^\alpha(\mathbf{x},\mathbf{x}').
  \label{eq:dc_approx}
\end{equation}
Critically, $\mathcal{D}^\alpha$ is \emph{never formed explicitly}:
the partition~\eqref{eq:dc_approx} is used only to derive assembly
formulas for each observable directly in terms of the local
eigenpairs $\{\lambda_n^\alpha,\psi_n^\alpha\}$, as detailed below.
Subsystem calculations are mutually independent for a fixed
$V_\text{eff}$, enabling fully parallel solution over all subsystems.

\subsubsection{Subsystem Solution via CheFSI}

Each subsystem KS problem is solved using Chebyshev-filtered
subspace iteration (CheFSI)~\cite{zhou2006self,zhou2006parallel},
which avoids explicit full diagonalization of $\mathcal{H}^\alpha$
while accurately spanning the occupied subspace.
Moreover, CheFSI incurs minimal orthogonalization and memory overhead, 
advantages that scale favorably with subsystem size~\cite{levitt2015}.
Given a block of $N_s^\alpha$ trial orbitals
$\Psi^\alpha \in \mathbb{R}^{N_d^\alpha \times N_s^\alpha}$,
each CheFSI iteration proceeds in four steps\cite{ghosh2017sparc}.

\noindent\textbf{(1) Chebyshev filtering.}
Apply the degree-$m$ Chebyshev polynomial
$p_m^\alpha(t) = C_m\!\left((t-c^\alpha)/e^\alpha\right)$
to the orbital block via the three-term recurrence,
requiring only $m$ sparse Hamiltonian-vector products:
\begin{equation}
  \Psi_f^\alpha = p_m^\alpha(\mathcal{H}^\alpha)\,\Psi^\alpha.
\end{equation}
The rapid growth of $C_m$ outside $[-1,1]$ suppresses all
eigencomponents above the cutoff $\lambda_c^\alpha$,
amplifying the occupied subspace while damping high-energy modes.

\noindent\textbf{(2) Projection.}
Project $\mathcal{H}^\alpha$ onto the filtered
basis $\Psi_f^\alpha$ to form the $N_s^\alpha\times N_s^\alpha$
reduced subspace matrices
\begin{equation}
  H_s^\alpha = (\Psi_f^\alpha)^T\mathcal{H}^\alpha\Psi_f^\alpha,
  \quad
  M_s^\alpha = (\Psi_f^\alpha)^T\Psi_f^\alpha.
\end{equation}

\noindent\textbf{(3) Subspace eigenproblem.}
Solve the $N_s^\alpha \times N_s^\alpha$ generalized eigenproblem
$H_s^\alpha y_n = \lambda_n^\alpha M_s^\alpha y_n$
\rev{via a dense generalized eigensolver}, yielding local eigenvalues
$\{\lambda_n^\alpha\}$ and subspace eigenvectors $\{y_n^\alpha\}$.
Because $N_s^\alpha \ll N_d^\alpha$, this dense solve is inexpensive.

\noindent\textbf{(4) Subspace rotation.}
Update the orbitals $\Psi^\alpha = \Psi_f^\alpha Y^\alpha$,
where $Y^\alpha = [y_1^\alpha,\ldots,y_{N_s^\alpha}^\alpha]$.
The resulting columns of $\Psi^\alpha$ are approximations to the
eigenvectors of $\mathcal{H}^\alpha$; together with
$\{\lambda_n^\alpha\}$ they form the complete output of CheFSI and
serve as input to the observable assembly described next.

The dominant cost per subsystem is
$\mathcal{O}(m\,N_d^\alpha\,N_s^\alpha)$ for Chebyshev filtering,
$\mathcal{O}(N_d^\alpha\,{N_s^\alpha}^2)$ for projection,
$\mathcal{O}({N_s^\alpha}^3)$ for the subspace eigenproblem,
and $\mathcal{O}(N_d^\alpha\,{N_s^\alpha}^2)$ for subspace rotation.
Since subsystem size is fixed as $N$ grows, each subsystem costs
$\mathcal{O}(1)$, yielding $\mathcal{O}(N)$ total cost and
$\mathcal{O}(N)$ memory.

\subsubsection{Assembly of Global Observables}

The electron density follows directly from the D\&C partition:
\begin{equation}
  \rho(\mathbf{x}) = \sum_\alpha p^\alpha(\mathbf{x})\,\tilde\rho^\alpha(\mathbf{x}),
  \label{eq:density}
\end{equation}
where $\tilde\rho^\alpha(\mathbf{x}) =
  2\sum_n g(\lambda_n^\alpha;\mu,\sigma)\,|\psi_n^\alpha(\mathbf{x})|^2$,
and the chemical potential $\mu$ is fixed by the charge-neutrality
constraint $\int_\Omega\rho\,d\mathbf{x} = N_e$.

All subsystem-resolved observables are expressed in terms of the
\emph{interior-region weight} $c_n^\alpha =
\int_{\Omega^\alpha_C}|\psi_n^\alpha(\mathbf{x})|^2\,d\mathbf{x}$,
which quantifies the contribution of local eigenstate $n$ to the
globally non-overlapping partition.
The band energy and entropy energy are:
\begin{align}
  E_\text{band} &= 2\sum_\alpha\sum_n
    g(\lambda_n^\alpha;\mu,\sigma)\,c_n^\alpha\,\lambda_n^\alpha, \\
  E_\text{entropy} &= -2k_BT\sum_\alpha\sum_n
    c_n^\alpha\,s(\lambda_n^\alpha;\mu,\sigma),
\end{align}
where $s(\lambda;\mu,\sigma) = g\log g + (1-g)\log(1-g)$.
Atomic forces are computed via a Hellmann-Feynman force
formulation~\cite{ghosh2017sparc}.
The electronic density of states (DOS) is constructed directly from
the local eigenpairs as
\begin{equation}
  D(E) =
  2\sum_{\alpha}\sum_n
  c_n^\alpha
  \delta\!\left(E-\lambda_n^\alpha\right).
  \label{eq:dos}
\end{equation}
The projected density of states (PDOS) is evaluated by projecting the
local KS orbitals onto atom-centered atomic orbitals.
For an atom $I$ with position $\mathbf{R}_I\in\Omega_C^\alpha$,
let $\phi_{I,lm}$ denote an atomic orbital centered at $\mathbf{R}_I$.
The projection weight of the $n$th local KS orbital is
\begin{equation}
  P_{n,I,lm}^\alpha
  =
  \left|
    \int
    \phi_{I,lm}^{*}(\mathbf{x})
    \psi_n^\alpha(\mathbf{x})
    \,d\mathbf{x}
  \right|^2 .
  \label{eq:pdos_projection}
\end{equation}
The corresponding orbital-resolved PDOS is
\begin{equation}
  D_{I,lm}(E)
  =
  \sum_n
  P_{n,I,lm}^\alpha
  \delta\!\left(E-\lambda_n^\alpha\right).
  \label{eq:pdos}
\end{equation}
The atom-resolved PDOS is obtained by summing over the atomic orbitals
associated with atom $I$,
\begin{equation}
  D_I(E)
  =
  \sum_{l,m} D_{I,lm}(E),
\end{equation}
and the species-resolved PDOS is obtained by summing over all atoms of
the corresponding chemical species.

Since the local eigenvalues and KS orbitals are already available from
the CheFSI solution, the DOS requires only the existing interior-region
weights, while the PDOS requires only local projections onto a fixed
number of atom-centered orbitals. Therefore, both DOS and PDOS retain
the overall $\mathcal{O}(N)$ complexity of XLSDFT.

\subsubsection{Self-Consistent Field Procedure}

The complete SCF cycle is depicted in Fig.~\ref{fig:xlsdft-overview}.
Each iteration comprises the following phases.

\noindent\textbf{Potential construction.}
The global effective potential
$V_\text{eff} = V_\mathrm{xc}[\rho] + \phi + V_\mathrm{nl}$
is assembled from the current electron density $\rho$.
The electrostatic component $\phi$ is obtained by solving
Eq.~\eqref{eq:poisson} via structured AMG preconditioned
CG (Sec.~\ref{subsec:poisson}), at
$\mathcal{O}(N)$ cost; the exchange-correlation and nonlocal
pseudopotential contributions are evaluated at $\mathcal{O}(N)$.

\noindent\textbf{Restriction of effective potentials.}
The global $V_\text{eff}$ is restricted to each subsystem domain
$\Omega^\alpha$, defining the local Hamiltonians
$\{\mathcal{H}^\alpha\}$. This step requires $\mathcal{O}(N)$
operations with local halo-exchange communication between
neighboring processes.

\noindent\textbf{Parallel CheFSI.}
Each process independently solves its assigned subsystem KS problems
via CheFSI (Chebyshev filtering, Rayleigh-Ritz projection, subspace
eigenproblem, and subspace rotation), as detailed in the preceding
subsection. No MPI communication occurs during this phase; the
dominant cost is $\mathcal{O}\!\left(\sum_\alpha m N_d^\alpha
N_s^\alpha\right)$, which is $\mathcal{O}(N)$ in aggregate.

\noindent\textbf{Chemical potential.}
The global chemical potential $\mu$ is determined by enforcing
charge neutrality, $\int_\Omega\rho\,d\mathbf{x}=N_e$, 
\rev{through
$N_e(\mu)=2\sum_\alpha\sum_n
g(\lambda_n^\alpha;\mu,\sigma)c_n^\alpha$.
Although this scalar equation can be readily solved using Brent's
method~\cite{press2007numerical}, each function evaluation requires an
\texttt{MPI\_Allreduce} over the distributed spectrum, making the
computation expensive at extreme scale.
We therefore employ a two-stage communication-efficient solver:
a global density-of-states histogram is first constructed with one
\texttt{MPI\_Allreduce} to obtain an approximate $\mu$, followed by
three to four Newton refinements of the exact occupation sum, each
requiring one collective for both $N_e(\mu)$ and its derivative.
Together with moving the computation of the interior weights
$c_n^\alpha$ from DDR to HBM, this reduces the number of global
reductions from approximately 25 to four or five and the overall
chemical-potential wall time from approximately 5--6~s to 80~ms.}

\noindent\textbf{Electron density update.}
With $\mu$ determined, the new electron density is assembled from
the local eigenpairs via Eq.~\eqref{eq:density}, at $\mathcal{O}(N)$
cost with no inter-process communication.

\noindent\textbf{Initial density and mixing.}
\rev{For the heterogeneous Li/LGPS system, we augment the conventional
superposition-of-atomic-densities initial guess with a material-specific
correction. The incoming and outgoing densities are then mixed using Periodic
Pulay~\cite{periodic_pulay_ref} with a material-masked real-space elliptic
preconditioner~\cite{lin_yang_elliptic,kumar_realspace_preconditioning}. For a
density residual $R_k$, the preconditioned residual $\widetilde{R}_k$ satisfies}
\begin{equation}
  \rev{\left[-\nabla^2+\kappa^2(\mathbf{x})\right]
  \widetilde{R}_k(\mathbf{x})=-\nabla^2R_k(\mathbf{x}).}
  \label{eq:elliptic-preconditioner}
\end{equation}
\rev{For homogeneous Si, $\kappa^2$ is constant. For Li/LGPS, we use
$\kappa^2(\mathbf{x})=s(\mathbf{x})k_{\mathrm{TF,Li}}^2$, where the smooth
material mask $s$ activates metallic screening in the Li regions and suppresses
it in insulating LGPS. This material-aware treatment, together with the
improved initial density, enables robust SCF convergence for the full
11-million-atom interface.}

\rev{\noindent Collective communication in the SCF cycle is concentrated in
the $\mathcal{O}(N)$ Poisson solve and the four to five reductions used by the
histogram-assisted Newton solve for $\mu$. The subsystem CheFSI computations
require no algorithmic inter-process communication.}

\subsection{HPC Implementation on the LineShine system}

\rev{Within each MPI process, XLSDFT combines OpenMP threading over subsystem
orbital blocks, hand-tuned ARM SVE vectorization for bandwidth-limited
finite-difference operations, and ARM SME matrix computation in custom dense linear-algebra kernels for
CheFSI projection and rotation. For the production runs reported here, all
required BLAS/LAPACK operations use custom implementations specialized to
XLSDFT's matrix dimensions, data layouts, and operation patterns. The general
implementation retains vendor-optimized BLAS/LAPACK outside these specialized
paths. Compared with the vendor-library path, this configuration recovers
approximately 1~GB of HBM per NUMA domain and avoids library-managed temporary
buffers in DDR. This additional capacity allows the finite-difference stencil,
nonlocal-projector application, Poisson solve, and most associated data to
reside in HBM.}
The following subsubsections describe each innovation in detail.

\subsubsection{Structured AMG Preconditioned Poisson Solver}
\label{subsec:poisson}

The global electrostatic Poisson system~\eqref{eq:poisson} is
discretized as the linear system $\mathbf{L}\mathbf{u} = \mathbf{f}$,
where $\mathbf{L}$ is the $2n_0$-order FD Laplacian with homogeneous
Dirichlet or periodic boundary conditions, and solved iteratively via
the conjugate gradient (CG) method.
Without preconditioning, the iteration count grows proportionally
to the grid diameter, exceeding several thousand iterations for systems
of tens of millions of atoms and rendering the Poisson solve intractable
within any practical time budget.
We accelerate convergence using a structured algebraic multigrid (AMG)
preconditioner based on the \mbox{StructMG}
framework~\cite{posterstructmg,structmg,semistructmg,fp16}.

\rev{To make the preconditioner efficient at scale, we retain the full 12th-order
operator in the outer CG solve but use a deliberately inexact seven-point
(3D7) approximation on the finest AMG level. This finest-level operator is
applied matrix-free, avoiding explicit storage of the dominant fine-grid
matrix and reducing memory traffic. The AMG V-cycle runs in FP32, while the
outer CG solver remains in FP64; this mixed-precision inexact preconditioning
reduces memory traffic and halo-exchange volume with only a negligible increase
in the V-cycle count. We further perform halo exchange in three directional
stages, successively along $x$, $y$, and $z$, forwarding edge- and corner-halo
data in subsequent stages. Each process therefore communicates directly with
six face neighbors rather than all 26 neighboring processes. Together with the
remaining application-level optimizations, this design preserves robust
convergence at scale, keeping the end-to-end Poisson-solve wall time below
$3$~s per SCF iteration for systems up to 200 million atoms.}

\subsubsection{SVE-Optimized Finite-Difference Stencil}
\label{subsec:stencil}

A dominant runtime kernel in XLSDFT is the
Hamiltonian-vector product $\mathcal{H}^\alpha\Psi^\alpha$, applied
$m$ times per CheFSI iteration.
For a $2n_0$-order 3D stencil at $n_0=6$, each grid point accumulates
$3\times2n_0=36$ neighbor contributions with fixed coefficients
$\{w_p\}_{p=0}^{n_0}$, yielding an arithmetic intensity of
$\approx0.97$~flop/byte at FP64 -- firmly memory-bandwidth-bound on LX2.
Optimization is guided by \texttt{objdump} disassembly of load-instruction
counts and placement.

\rev{\textbf{Custom data layouts} reorganize $\Psi^\alpha$ into
16-band tiles for the stencil and Chebyshev recurrence,
and into thread-partitioned band--grid packs for projection; 
the hot-path kernels consume these
layouts directly, so filter and $\mathcal{H}\Psi$ steps avoid per-iteration
layout conversion.}
\rev{\textbf{Threading and vectorization} partition band tiles and grid points
across OpenMP threads with no cross-band dependences; within each thread,
ARM SVE updates 16~FP64 grid coefficients per stencil point.}
\rev{\textbf{Multi-level cache blocking} tiles the $(i,j,k)$ grid in
$64\times4\times8$ chunks over the packed input, improving
reuse on bandwidth-bound updates; non-temporal stores write
$\mathcal{H}^\alpha\Psi^\alpha$ back in the same packed form, or fused
layouts for projection during $\mathcal{H}\Psi$.}
\rev{\textbf{Axis neighbor reuse} applies boundary-aware stencil updates along
$x$, $y$, and $z$ on 16-wide packed vectors, reusing shifted neighbor lanes
where possible and interleaving loads with multiply-accumulates to hide
LX2 load-use latency.}
\rev{\textbf{Nonlocal pseudopotential} ($V_{\mathrm{nl}}$) groups projectors by
atom; forward projection and $\gamma$ scaling accumulate 16 bands per SVE
inner loop, and backprojection uses a prebuilt reverse-CSR over grid points
so each output location gathers its contributions without atomics or shared
workspace.}
\rev{\textbf{Chebyshev recurrence} fuses the filter AXPY on the tiled
layouts.}
\textbf{FP16 orbital warm-start} stores orbital guesses from the previous
SCF iteration at half precision, halving HBM traffic for the first
Chebyshev filter step while all subsequent arithmetic uses FP64.

Together, these optimizations improve FD stencil throughput from
1.66 Tflop/s to 5.25 Tflop/s per node, a $3.2\times$ speedup over a plain SVE
implementation.
\rev{On the 20480-node 100-million-atom calculation, the optimized Laplacian and
$V_{\mathrm{nl}}$ kernels reach aggregate HBM bandwidth peaks of
143~PB/s (89\% of peak) and 107~PB/s, respectively.}

\subsubsection{Custom Dense Linear Algebra}
\label{subsec:sme}

The Rayleigh-Ritz projection and subspace rotation in CheFSI reduce
to dense matrix-matrix products:
\begin{align*}
  H_s^\alpha &= (\Psi_f^\alpha)^T(\mathcal{H}^\alpha\Psi_f^\alpha),
  \quad M_s^\alpha = (\Psi_f^\alpha)^T\Psi_f^\alpha
    \in\mathbb{R}^{N_s^\alpha\times N_s^\alpha},\\
  \Psi^\alpha & = \Psi_f^\alpha Y^\alpha,
  \quad (N_d^\alpha\times N_s^\alpha)\times(N_s^\alpha\times N_s^\alpha).
\end{align*}
At the subsystem sizes of interest
($N_d^\alpha\approx 200{,}000$, $N_s^\alpha\approx 640$), we execute these
products with custom FP64 DGEMM and DSYRK kernels \rev{on the packed layouts},
together with specialized implementations of the remaining BLAS/LAPACK
operations required by these runs.
\rev{Each kernel uses SME outer-product accumulation with multi-level blocking;
projection GEMM and SYRK
partition the $K$ dimension using the Stream-K algorithm, keeping per-thread
partial tiles that are reduced at the end through SVE, and issue software
gather-based prefetch (\texttt{svprfw\_gather\_index}) on upcoming $A$/$B$
panels.
Subspace rotation uses a corresponding tile-aware SME GEMM.
The generalized eigenproblem is solved by a specialized DSYGVD implementation:
contiguous SVE dot/norm/axpy in Cholesky
factorization and Householder tridiagonalization, panel-wise reduction,
and a specialized parallel implicit-QL sweep over eigenvector
rows.}
For the 100-million-atom calculation, the projection and subspace-rotation
phases reach \rev{1.20 and 1.31~Eflop/s}, respectively.


\section{How Performance Was Measured}
\label{sec:measurement}

\subsection{Applications Used to Measure Performance}

XLSDFT is a new C++ implementation of the 
linear-scaling KS-DFT framework described in Sec.~\ref{sec:innovations},
parallelized with MPI, OpenMP, and ARM SVE/SME intrinsics.
\rev{The source code is publicly available at
\url{https://github.com/NSCCSZ-HPC/XLSDFT}.}
All reported calculations use FP64 throughout, with two exceptions:
orbital initial guesses are stored in FP16 between SCF iterations to
reduce memory bandwidth, and the AMG preconditioner for the CG Poisson
solver runs in FP32 with an FP64 outer iteration
(Sec.~\ref{subsec:poisson}).
The code is compiled with a vendor-optimized \texttt{clang} 17.0.6 at
\texttt{-O3 -march=armv9-a+sve+sme}.
\rev{All calculations use an SCF convergence tolerance of
$5\times10^{-4}$ for the relative electron-density residual.}

Two applications are used to measure performance:

\noindent\textbf{Silicon crystal.}
\rev{We used a diamond-cubic silicon crystal with 100~million atoms as
the endpoint of the systematic weak-scaling study. An additional
200-million-atom crystal successfully converged on the same 20,480
nodes as an extreme-scale capability demonstration.}
This system provides a well-controlled, single-element reference with known
convergence properties, enabling direct comparison to prior
$\mathcal{O}(N)$ DFT implementations.
GGA norm-conserving pseudopotential~\cite{hamann2013optimized};
grid spacing $h \approx 0.5$~Bohr; $12$th FD order ($n_0 = 6$);
Chebyshev filter degree $m = 2\rev{4}$;
buffer radius $R_\text{buffer} \approx 9.2 $~Bohr.

\noindent\textbf{Li/LGPS solid-state battery interface.}
An 11-million-atom Li/LGPS interface cell ($57\times44\times90$~nm)
is used to demonstrate science at experimentally relevant scale.
The atomic structure was generated by MLFF-MD (Stage~1 of the workflow
in Sec.~\ref{sec:overview}) and thermally equilibrated at 300~K.
GGA norm-conserving pseudopotential~\cite{hamann2013optimized};
grid spacing $h \approx 0.4$~Bohr; $12$th order FD ($n_0 = 6$);
$m = 2\rev{4}$; $R_\text{buffer} = 8.0$~Bohr.
\rev{The production calculation reached the SCF convergence threshold after
167 accumulated iterations. Robust convergence was enabled by the improved SAD-based initial density
and the material-masked elliptic preconditioner described in
Sec.~\ref{sec:innovations}.}

\rev{XPS depth profiling used a PHI VersaProbe 4 with an Ar$^+$ sputtering
source. After reaction, the LGPS pellet was separated from Li metal and
analyzed in situ from the reacted surface. Spectra acquired after 0, 20, and
50~min of sputtering represent the interface, sub-interface, and bulk-like
LGPS regions, respectively.}

\subsection{HPC System}

The \emph{LineShine} supercomputer~\rev{\cite{lu2026lineshine}} is an
exascale system \rev{with over 20,000 compute nodes,} developed by the
National Supercomputing Center in Shenzhen (NSCC-SZ), China.
Each node contains two Armv9-based LX2 processors.
The LX2 integrates two compute dies (304 cores total) and eight
on-package HBM stacks (32~GB, 4~TB/s aggregate bandwidth) in a single
package.
Each compute die contains 152 cores and is paired with 128~GB of
off-package DDR memory organized into four NUMA domains (eight per
processor, sixteen per node), for a total of 256~GB DDR per processor.
Within each NUMA domain, cores share HBM.
The LX2 supports FP64, FP32, FP16, and INT8 through SME and SVE units,
delivering up to 60.3~Tflop/s FP64.
The compute nodes in the LineShine system are interconnected via
a LingQi high-speed network, with a dual-plane, multi-rail fat-tree
topology, providing 1.6~Tb/s bandwidth per node.
The entire system delivers over 2.5~Eflop/s peak performance in FP64.

\begin{figure}[htb]
  \centering
  \includegraphics[width=0.9\columnwidth]{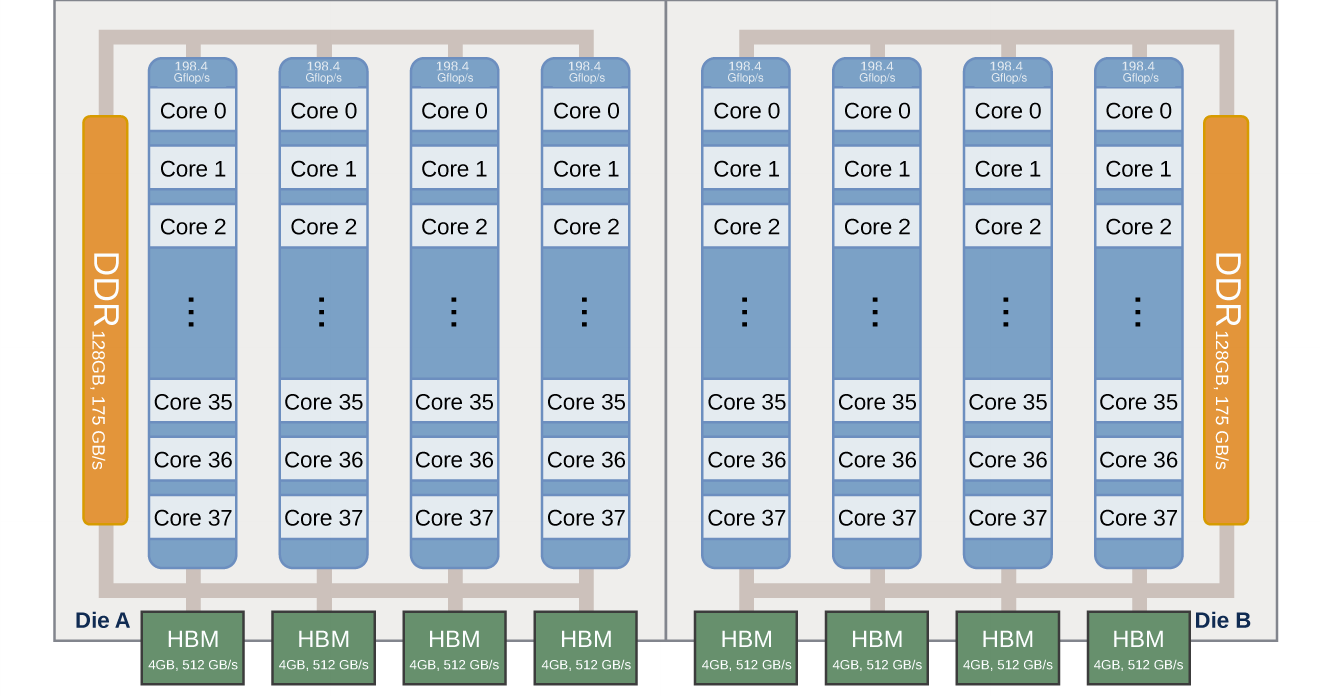}
  \caption{Architecture of the LX2 processor: two compute dies
    (304 cores total), eight on-package HBM stacks in a single package. 
    }
  \label{fig:lx2}
\end{figure}

\subsection{Measurement Methodology}

Wall time is measured with \texttt{MPI\_Wtime}, bracketed by
\texttt{MPI\_Barrier} calls at the entry and exit of each timed region.
\rev{Reported times are averaged over all but the first and last SCF iterations.}
I/O (reading the atomic structure, writing observables) is included
in the time-to-solution wall time but excluded from the sustained
Pflop/s calculation.

\rev{Sustained FP64 performance (Pflop/s) is computed from the analytic
operation count and measured CheFSI wall time, rather than total SCF wall time.}
The flop counts are estimated analytically for the two dominant kernel
\rev{groups}: the FD stencil ($2 \times 6n_0 \times 3$ flop per grid point per
orbital per application, summed over all $m$ filter steps and all
subsystems) and \rev{the DGEMM and DSYRK routines in CheFSI projection
($2N_d^\alpha {N_s^\alpha}^2$ and $N_d^\alpha {N_s^\alpha}^2$ flop
per call, respectively); and the DGEMM in subspace rotation
($2N_d^\alpha {N_s^\alpha}^2$ flop per call). The projection and
subspace-rotation rates are computed from these counts and their corresponding
wall times.}


\section{Performance Results}
\label{sec:results}

\begin{figure*}[t]
  \centering
  \includegraphics[width=0.92\textwidth]{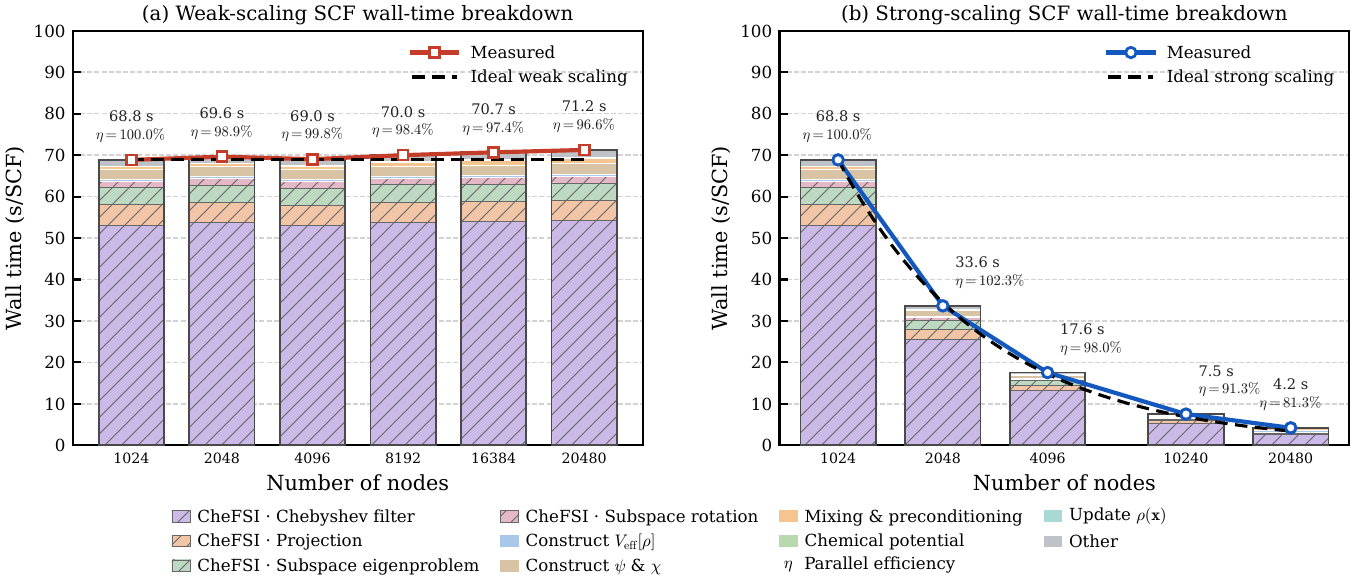}
  \caption{Weak (left) and strong (right) scalability \rev{and SCF wall-time
    breakdown} of XLSDFT on the silicon crystal on LineShine.
    \textit{Left:} atom count and node count scaled proportionally from
    \rev{$5,242,880$ atoms on 1,024~nodes} to 104,857,600~atoms on 20,480~nodes
    ($5{,}120$~atoms/node);
    efficiency referenced to \rev{1,024~nodes}.
    \textit{Right:} fixed 5,242,880-atom silicon crystal scaled
    from 1,024 to 20,480~nodes;
    efficiency referenced to 1,024~nodes.
    \rev{Stacked bars show component wall times, solid lines show total
    measured wall time, and dashed lines show ideal scaling.}
    \rev{Endpoint efficiencies are 96.6\% for weak scaling and 81.3\% for
    strong scaling. Labels above each point report total SCF wall time and
    parallel efficiency.}
    }
  \label{fig:si-scaling}
\end{figure*}

\subsection{Scalability}

Fig.~\ref{fig:si-scaling} presents the weak and strong scalability
\rev{and corresponding SCF wall-time breakdowns} of XLSDFT on the silicon
crystal.

\noindent\textbf{Weak scalability.}
XLSDFT scales from \rev{5,242,880 atoms on 1,024~nodes} to 104,857,600~atoms
on 20,480~nodes, a \rev{20-fold} increase in both atom count
and node count.
Weak parallel efficiency reaches \rev{96.6\%} at 20,480~nodes, with
\rev{71.2~s} per SCF iteration at the 100-million-atom endpoint.
\rev{This near-ideal weak scaling is enabled in part by the
histogram-assisted Newton chemical-potential solver, which reduces the
number of global reductions from approximately 25 to four or five and lowers
this step from approximately 5--6~s to 80~ms. Improvements to the
finite-difference and dense-matrix kernels also mitigate rank-to-rank
compute-time variation and thereby reduce waiting at synchronization points.}

\noindent\textbf{Strong scalability.}
For a fixed 5,242,880-atom silicon crystal, XLSDFT achieves 
\rev{81.3\% efficiency across the full 20-fold increase from 1,024 to
20,480~nodes,
reducing wall time per SCF iteration from $\sim$68.8~s to
$\sim$4.2~s.
Efficiency degradation at high node counts reflects the transition to the
communication-dominated regime as the per-node subsystem grain shrinks.}

\noindent\textbf{Memory scaling.}
\rev{The total memory footprint scales linearly with atom count, consistent
with the $\mathcal{O}(N)$ memory design. For the reported runs, using specialized
dense-linear-algebra implementations instead of the vendor BLAS/LAPACK path
recovers approximately 1~GB of HBM per NUMA domain and eliminates associated
temporary DDR buffers. The recovered HBM capacity and the migration of major
memory-intensive operations to HBM help enable the 200-million-atom calculation
on the same 20,480 nodes used for the 100-million-atom case.}

\subsection{Sustained Performance}

\rev{At 104,857,600~atoms on 20,480~LineShine nodes, XLSDFT sustains
157.9~Pflop/s in FP64 during CheFSI. The custom dense kernels reach
1.20~Eflop/s during projection and 1.31~Eflop/s
during subspace rotation. Conversely, the $12$th-order finite-difference
Hamiltonian-vector product has an arithmetic intensity of approximately
0.97~flop/byte and is memory-bandwidth-bound. Peak/sustained per-node HBM
bandwidths are 7.91/4.01~TB/s for Si and 5.74/0.186~TB/s for Li/LGPS.
The peak is the maximum 0.1-second sample across kernels, whereas the sustained
value is the median 1.0-second CheFSI sample. Extrapolating these per-node rates
over 20,480 nodes gives estimated aggregate peak/sustained bandwidths of
161.9/82.1~PB/s for the Si calculations and 117.6/3.80~PB/s for Li/LGPS.
I/O is excluded from the performance rates as described in
Sec.~\ref{sec:measurement}.}

\subsection{Time-to-Solution}

\rev{Table~\ref{tab:time-to-solution} compares time to solution and FP64
performance for the three calculations. Total wall time includes initialization,
I/O, and energy evaluation for all systems, plus PDOS calculation for Li/LGPS.}

\begin{table}[htb]
  \centering
  \footnotesize
  \setlength{\tabcolsep}{3pt}
  \caption{\rev{Wall time and FP64 performance of the largest XLSDFT
  calculations on LineShine.}}
  \label{tab:time-to-solution}
  \begin{tabular}{@{}lrrr@{}}
    \toprule
    Metric & Si (100M) & \rev{Si (200M)} & \rev{Li/LGPS (11M)} \\
    \midrule
    \rev{SCF iterations}             & \rev{7}     & \rev{7}      & \rev{167} \\
    \rev{Total wall time (s)}        & \rev{1082.7} & \rev{2280.4} & \rev{6936.2} \\
    Wall time (s/SCF)            & \rev{71.2}  & \rev{143.7}  & \rev{36.4} \\
    \rev{Sustained CheFSI (Pflop/s)} & \rev{157.9} & \rev{157.6} & \rev{96.8} \\
    \rev{Projection (Eflop/s)}   & \rev{1.20}  & \rev{1.20}   & \rev{0.10} \\
    \rev{Rotation (Eflop/s)}     & \rev{1.31}  & \rev{1.28}   & \rev{0.57} \\
    \bottomrule
  \end{tabular}
\end{table}

\subsection{Scientific Results: Li/LGPS Battery Interface}

\begin{figure*}[t]
  \centering
  \includegraphics[
    width=0.90\textwidth,
    trim=30bp 80bp 60bp 95bp,
    clip
  ]{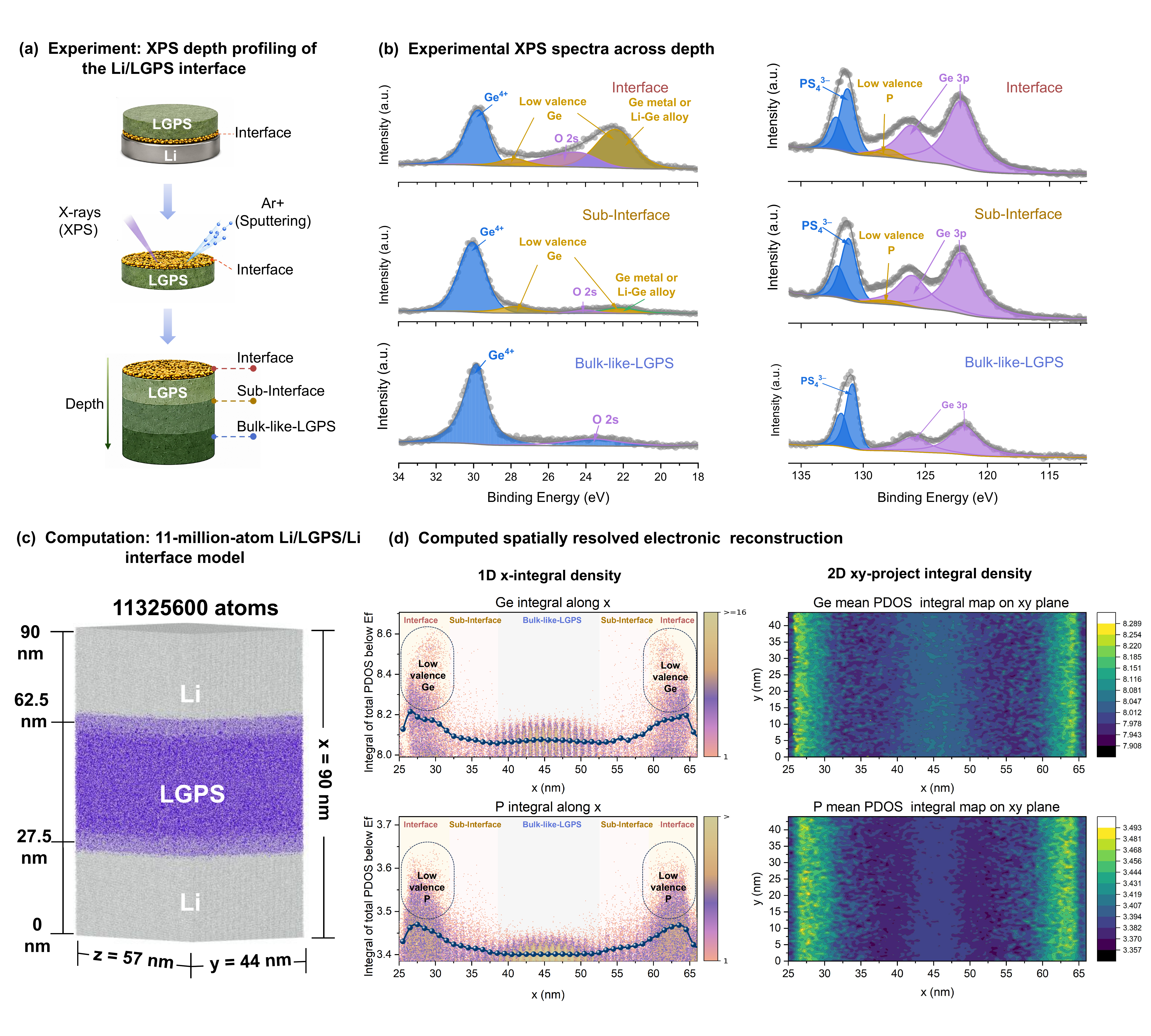}
  \caption{\rev{Experimental and computational characterization of spatially
    heterogeneous low-valence-state formation at the Li/LGPS interface.
    (a)~XPS depth-profiling schematic.
    (b)~Depth-dependent Ge and P XPS spectra from interfacial, sub-interface,
    and bulk-like LGPS regions.
    (c)~The 11,325,600-atom Li/LGPS/Li model, approximately
    $90\times44\times57$~nm.
    (d)~Computed interface-normal distributions and $xy$-plane projections of
    occupied-PDOS integrals for Ge and P. Larger integrals indicate lower local
    valence; both representations show low-valence Ge and P enriched near the
    interfaces and spatially heterogeneous in-plane.}}
  \label{fig:battery-results}
\end{figure*}

To overcome the system-size limitations of conventional first-principles
interfacial studies, we constructed an 11-million-atom Li/LGPS interfacial
model at experimentally relevant dimensions (Fig.~\ref{fig:battery-results}c).
The model spans $57\times44\times90$~nm, far beyond conventional idealized
small-interface models, bringing the computational length scale into the range
accessible to experimental observation.
\rev{The full calculation reached SCF convergence. Convergence of this
heterogeneous metallic--insulating
system was enabled by the improved SAD-based initial density and the
material-masked elliptic preconditioner combined with Periodic Pulay mixing.}
\rev{Its performance is summarized in Table~\ref{tab:time-to-solution}.}

\rev{Atom-resolved valence-state analysis integrates occupied PDOS below the
Fermi level for Ge and P, with larger integrals indicating lower local valence.
The interface-normal distributions and $xy$-plane projections show enhanced
integrals near both interfaces that decay toward bulk-like LGPS, together with
pronounced in-plane heterogeneity (Fig.~\ref{fig:battery-results}d). Thus,
reduction is localized near the interfaces rather than spatially uniform.

XPS depth profiling independently shows low-valence Ge, Ge metal or Li--Ge
alloy, and low-valence P concentrated in the interfacial region and diminishing
with sputtering depth toward bulk-like LGPS
(Fig.~\ref{fig:battery-results}a,b). The consistent interface-to-bulk trend
validates the XLSDFT electronic reconstruction, while the calculation
additionally resolves in-plane heterogeneity that is difficult to access
experimentally.}

\section{Implications}
\label{sec:implications}

XLSDFT removes a length-scale barrier that has constrained 
\emph{ab initio} materials simulation for decades.
By reducing KS-DFT scaling from $\mathcal{O}(N^3)$ to $\mathcal{O}(N)$
\rev{and sustaining 157.9~Pflop/s in FP64 for the systematic 100-million-atom
study, with custom dense kernels reaching 1.31~Eflop/s in a critical
subphase,} it makes production-quality electronic-structure calculations
routine at scales previously accessible only to empirical or
machine-learning approaches---and then only for atomic forces, not
electronic-structure observables.
The \rev{100-million-atom silicon scaling study, the 200-million-atom silicon
capability demonstration,} and the DFT simulation
of a physically representative 11-million-atom all-solid-state battery
interface establish that this capability is not a one-off feat but a
deployable tool for real materials problems.

For the Li/LGPS interface, the implications are direct.
Prior DFT was confined to a few hundred atoms---too small to resolve
the nanoscale interphase that governs battery degradation.
\rev{The computed PDOS analysis shows that low-valence Ge and P are enriched
near the Li/LGPS interfaces, decay toward bulk-like LGPS, and vary substantially
within the interfacial plane (Fig.~\ref{fig:battery-results}d). XPS depth
profiles independently show the same interface-to-bulk reduction trend
(Fig.~\ref{fig:battery-results}a,b), while XLSDFT additionally resolves
in-plane heterogeneity difficult to access experimentally. Together, these
experimental and computational results identify concrete design targets:
suppress interfacial reduction and block electron leakage.}
Rational interface engineering can now be guided by first-principles
electronic structure rather than empirical trial and error, at
physically representative length scales for the first time.
More broadly, the divide-and-conquer, CheFSI, and exascale-optimized
implementation strategy \rev{is applicable to a broad class of systems} requiring orbital-resolved
observables at the million-atom scale with mixed metallic and insulating
character---oxide and halide solid-state electrolytes, cathode/electrolyte
contacts, semiconductor device interfaces, complex oxide heterostructures,
and radiation damage in structural materials.
\rev{The multiscale MLFF--DFT workflow, coupled to experimental validation,
is transferable to other reactive solid--solid interfaces and provides a
foundation for future closed-loop digital--physical materials discovery.}
The demonstration that $\mathcal{O}(N)$ KS-DFT can operate at extreme
scale on an exascale supercomputer opens a new frontier in which the physically relevant and the
computationally accessible length scales finally coincide.


\section*{Acknowledgment}
This work was supported by the Guangdong Science and Technology Program
under Grant No.~2024B0101040005, the National Natural Science Foundation
of China under Grant No.~12404264, and the Greater Bay Area Institute of
HPC-AI Co-Driven Innovation. Computations were performed on the LineShine
supercomputer at the National Supercomputing Center in Shenzhen (NSCC-SZ).
OpenAI Codex was used to assist with language editing and LaTeX preparation
throughout the manuscript. All scientific content, data, and conclusions were
reviewed and verified by the authors.


\bibliographystyle{IEEEtran}
\bibliography{bib/references}

\end{document}